\documentclass[aps,prmaterials,reprint,superscriptaddress]{revtex4-1}
 \pdfoutput=1
\usepackage{graphicx,charter}

\begin{document}

\title{Structural evolution and skyrmionic phase diagram of the lacunar spinel GaMo$_4$Se$_8$}

\author{Emily C. Schueller}
\author{Daniil A. Kitchaev}
\author{Julia L. Zuo}
\author{Joshua D. Bocarsly}
\author{Joya A. Cooley}
\author{Anton Van der Ven}
\author{Stephen D. Wilson}
\author{Ram Seshadri}
\affiliation{Materials Department and Materials Research Laboratory\\
University of California, Santa Barbara, Santa Barbara, CA, 93106}

\date{\today}

\begin{abstract}
In the $AB_4Q_8$ lacunar spinels, the electronic structure is described on the basis of inter- and intra-cluster 
interactions of tetrahedral $B_4$ clusters, and tuning these can lead to myriad fascinating electronic and magnetic
ground states. In this work, we employ magnetic measurements, synchrotron X-ray and neutron scattering, and 
first-principles electronic structure calculations to examine the coupling between structural and magnetic phase 
evolution in GaMo$_4$Se$_8$, including the emergence of a skyrmionic regime in the magnetic phase diagram. 
We show that the competition between two distinct Jahn-Teller distortions of the room temperature cubic 
$F\overline{4}3m$ structure leads to the coexistence of the ground state $R3m$ phase and a metastable $Imm2$ phase. 
The magnetic properties of these two phases are computationally shown to be very different, with the $Imm2$ phase 
exhibiting uniaxial ferromagnetism and the $R3m$ phase hosting a complex magnetic phase diagram including equilibrium 
N{\'e}el--type skyrmions stable from nearly $T$\,=\,28\,K down to $T$\,=\,2\,K, the lowest measured temperature. 
The large change in magnetic behavior induced by a small structural distortion reveals that GaMo$_4$Se$_8$ 
is an exciting candidate material for tuning unconventional magnetic properties \textit{via} mechanical means.

\end{abstract}


\maketitle

\section{Introduction}
In recent years, there has been an increasing recognition that in some transition metal compounds, metal--metal 
bonding can often interact with electron correlation resulting in unconventional magnetic and electronic behavior, 
as well as unusually strong coupling between magnetic, electronic, and structural degrees of freedom \cite{Mazin2012,Hiroi2015,Streltsov2016,Schueller2019}. An important class of materials where this becomes 
evident are the lacunar spinels, which are characterized by tetrahedral clusters of transition metal atoms \cite{Kim2014}. Two members of the lacunar spinel family, GaV$_4$S$_8$ and GaV$_4$Se$_8$, have garnered significant attention in recent years as multiferroic materials as well as N\'eel--type skyrmion hosts \cite{Fujima2017,Ruff2017,Bordacs2017}.  At low temperatures, these materials crystallize in the polar $R3m$ space group \cite{Pocha2000} and support complex magnetic phase diagrams with an ordered moment of 1\,$\mu_B$/V$_4$ cluster.  The related Mo--containing compounds, GaMo$_4$S$_8$ and GaMo$_4$Se$_8$, have been 
less studied, although they are reported to have a similar low temperature crystal structure, but with a compressive rather than elongative distortion along the cubic $[111]$  direction \cite{Pocha2000,Francois1992,Francois1990}.

Bulk magnetic measurements have revealed similar metamagnetic  behavior between the Mo and the V--containing compounds \cite{Rastogi1983,Rastogi1987}, while computational analysis of lacunar spinel ferromagnets suggests that all materials with this structure type are likely to host skyrmions across wide temperature ranges \cite{Kitchaev2020}. However, because Mo is a heavier element than V, the Mo--based compounds exhibit increased spin--orbit coupling, leading to more pronounced magnetic anomalies over greater regions of magnetic field and temperature. Additionally, spin--orbit coupling can affect the electronic  properties through splitting of degenerate molecular orbitals, as seen in GaTa$_4$Se$_8$ \cite{Jeong2017}.  A recent report on GaMo$_4$S$_8$ indicates that this material harbors complex modulated magnetic phases and that the increased spin--orbit coupling in the material leads to unusual ``waving" effects on the magnetic ordering vectors \cite{Butykai2019}.  

In this work, we perform a detailed theoretical and experimental study of GaMo$_4$Se$_8$ in order to characterize 
the coupling between the crystal structure evolution and magnetic properties.  We discover that at the structural 
phase transition, GaMo$_4$Se$_8$ can transform into two polar space groups, the reported rhombohedral $R3m$ phase, 
which is the ground state, and an orthorhombic $Imm2$ phase, which is metastable. On the basis of computational 
and experimental data, we show that the $R3m$ phase of GaMo$_4$Se$_8$ hosts a rich magnetic phase diagram with 
long wavelength magnetic order, including cycloids and N\'eel--type skyrmions, with periodicities on the order 
of 16\,nm. In contrast, the competing $Imm2$ phase is suggested from first-principles calculations to be a 
strongly uniaxial ferromagnet. The dramatic change in magnetic behavior associated with a subtle change in 
crystal structure highlights the strong coupling between crystal structure and magnetism in GaMo$_4$Se$_8$, and 
suggests an avenue for tuning skyrmion and other exotic magnetic phases \textit{via} strain.   

\section{Methods}

\subsection{Computational Methods}

All electronic structure calculations were performed using the Vienna Ab--Initio Simulation Package 
(VASP) \cite{Kresse1996} with projector-augmented-wave pseudopotentials \cite{Blochl1994,Kresse1999} within the 
Perdew--Burke--Ernzerhof generalized gradient approximation \cite{Perdew1996}. We do not employ an atom-centered
Hubbard--$U$ correction as our previous analysis of the related GaV$_4$Se$_8$ system showed that this leads to 
an incorrect high--moment magnetic configuration by penalizing metal--metal bonding \cite{Schueller2019}. 
Structural distortions from the high temperature cubic structure to the low temperature orthorhombic and 
rhombohedral structures were generated using ISODISTORT \cite{Campbell2006}. Structural stability calculations 
used a Gamma--centered $k$--point grid of 8$\times$8$\times$8 and an energy cutoff of 500\,eV with a collinear 
magnetic configuration of 1\,$\mu_B$/Mo$_4$.  Computational results were parsed and visualized with the python 
package pymatgen \cite{Ong2013}.  Noncollinear magnetic calculations accounting for spin--orbit coupling followed 
the methodology described by Kitchaev \textit{et al.} \cite{Kitchaev2020}. All calculations were performed statically, 
on the basis of the experimentally determined crystal structures, and converged to 10$^{-6}$\,eV in total energy. 
To reduce noise associated with $k$--point discretization error, all spin--configuration energies were referenced 
to the energy of a $c$-axis ferromagnet computed within the same supercell. Noncollinear magnetic configurations
were initialized by evenly distributing the magnetic moment of each Mo$_4$ tetrahedron on the four Mo atoms,
i.e. as a locally ferromagnetic configuration of 0.25 $\mu_B$ per Mo. The magnetic moment of each 
tetrahedron was then obtained from the DFT data by summing the individual magnetic moments projected onto the four Mo 
atoms.

The magnetic cluster expansion Hamiltonian \cite{Drautz2004, Thomas2017, VdV2018, Kitchaev2020} used to construct 
the magnetic phase diagram was derived by considering single--spin and nearest--neighbor pair interactions, as 
shown in Supplementary Figure S4 \cite{SI}. The interaction strengths were parametrized on the basis of collinear and 
spin--wave enumerations up to a supercell size of 4, as well as several sets of spin configurations iteratively 
generated to independently constrain the terms in the Hamiltonian.\cite{vandeWalle2002} Note that due to the 
very small energy scale of magnetocrystalline anisotropy, we fit the single-site anisotropy coefficients 
independently of the pair interactions, by fitting to the energy of rotating the ground-state collinear spin 
configurations within the primitive cell of the structure, so as to eliminate all noise arising from changes 
in the $k$--point grid \cite{Kitchaev2020}.  The resulting parametrization leads to an error below 
5\,$\mu$eV/Mo$_4$ for the magnetocrystalline anisotropy, and below 1\,meV/Mo$_4$ across all non--collinear 
spin configurations, as shown in Supplementary Figure S5 \cite{SI}. The full Hamiltonians and fitted interaction parameters 
for the $R3m$ and $Imm2$ phases are given in Supplementary Tables 1 and 2 \cite{SI}.

\subsection{Experimental Methods}

GaMo$_4$Se$_8$ powder was obtained by reaction of Ga pieces and ground Mo and Se powders with an approximately 
50\% excess of elemental Ga. The elements were reacted in an evacuated fused silica tube in 1\,g batches with 
a heating ramp rate of 8$^\circ$\,C/min to 1010$^\circ$\,C, held for 20 hours, and quenched \cite{Jakob2007}.  

High resolution synchrotron powder X-ray diffraction was performed at the Advanced Photon Source at Argonne 
National Laboratory at the 11-BM beamline.  Approximately 30\,mg of powder was loaded into a kapton capillary 
and placed into a cryostat. Temperature--dependent scans with a 10\,minute exposure time were taken from $T$\,=\,10\,K to $T$\,=\,250\,K. Rietveld refinements were performed using the Topas analysis package \cite{Coelho2018}.  Low--temperature structure solution was performed by first fitting the reported rhombohedral structure and then indexing the remaining structural peaks 
using the EXPO2014 software.  The symmetry of the unit cell generated by EXPO was compared to various subgroups 
of the high temperature $F\overline{4}3m$ structure using the ISODISTORT software package \cite{Campbell2006}.  
A structure for the corresponding orthorhombic $Imm2$ space group was generated using ISODISTORT and Rietveld 
refinements of both the rhombohedral and orthorhombic phases were performed using TOPAS Academic \cite{Coelho2018}. 
Due to $hkl$-dependent strain, the Stephens peak shape was used in conjunction with Gaussian size broadening to 
fit the highly anisotropic peak shapes. Crystal structures were visualized using the VESTA software 
suite \cite{Momma2011}.

Small angle neutron scattering measurements were performed at the National Institute of Standards and Technology 
(NIST) NG-7 neutron beamline.  Approximately 100\,mg of powder was cold-pressed into a 5\,mm pellet and placed 
into a cryostat. Exposures were taken over 8\,h durations at various applied magnetic fields with the field applied parallel to the neutron beam. Three scans were taken below the magnetic ordering temperature and a 3\,h background scan was performed with no applied field above the magnetic ordering temperature. The data integration and background subtraction was performed in the IGOR Pro software \cite{Kline2006}.

Bulk magnetic measurements were performed on a Quantum Design MPMS 3 SQUID Vibrating Sample Magnetometer, 
with approximately 30\,mg of powder loaded into a plastic container and mounted in a brass holder.  
DC magnetization \textit{vs.} field scans were performed on increasing field and AC susceptibility \textit{vs.} 
field measurements were performed on decreasing field. For magnetoentropic mapping measurements, temperature-dependent 
magnetization measurements from 2\,K to 40\,K were performed with applied magnetic fields varying from 
0\,Oe to 4000\,Oe in increments of 75\,Oe. Derivatives and $\Delta$S$_m$ calculations were performed using 
python code available from Bocarsly \textit{et al.} \cite{Bocarsly2018}.

\section{Results}

\begin{figure*}[t!]
\includegraphics[width=\textwidth]{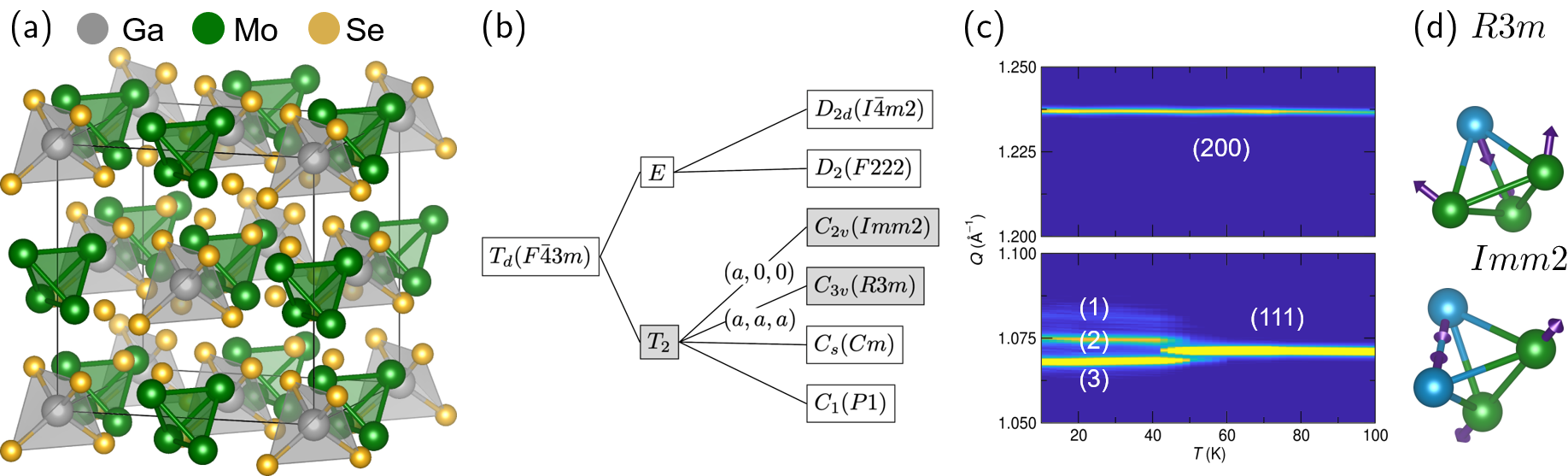}%
\caption{(a) The cubic $F\overline{4}3m$ crystal structure of GaMo$_4$Se$_8$, characterized by tetrahedral 
clusters of Mo atoms. (b) All subgroups of $F\overline{4}3m$ which keep the primitive unit cell volume the same. 
$R3m$ and $Imm2$ both follow the $T_2$ irrep but through different order parameter directions.  
(c) The evolution of peaks upon cooling in synchrotron XRD data. At 50\,K the (111) cubic peak splits into 
3 while the (200) cubic peak stays undistorted. Because the middle peak of the (111) split shifts relative to 
the peak in the cubic phase while the (200) peak does not shift, it is not possible to explain the low 
temperature structure by a combination of the cubic phase and a single low temperature phase.  
(d) The distortion of the Mo$_4$ tetrahedron as it goes into the $R3m$ and $Imm2$ space groups. 
Both are polar, but the $Imm2$ cluster has lower symmetry, with three unique bond lengths, compared to 
two in the $R3m$ cluster.}
\label{fig1}
\end{figure*} 
 
The room temperature crystal structure of GaMo$_4$Se$_8$ is highlighted in Figure \ref{fig1}(a), characterized by tetrahedral clusters of Mo atoms which behave like molecular units with a spin of 1\,$\mu_B$ per Mo$_4$ cluster.  Through experimental and computational techniques, we find that GaMo$_4$Se$_8$ follows qualitatively similar behavior to GaV$_4$Se$_8$, with complex magnetic order arising from a Jahn--Teller distortion of the high--temperature $F\overline{4}3m$ structure. However, in addition to the $R3m$ ground state, GaMo$_4$Se$_8$ forms a competing metastable $Imm2$ phase, defined by an orthorhombic Jahn--Teller distortion and distinctly different magnetic properties. We first characterize the formation and coexistence of these two structures, and then establish how the character of the low--temperature distortion controls the magnetic phase diagram of each phase.

Figure \ref{fig1}(c) shows high-resolution synchrotron powder diffraction data as a function of temperature. 
At $T$\,=\,50\,K, the (111) cubic peak splits into 3 peaks, with the two outer peaks corresponding to an 
$R3m$ distortion. Previous X--ray diffraction studies of the structural phase transition of GaMo$_4$Se$_8$ have 
rationalized the presence of extra peaks below the transition temperature ($T$\,=\,51\,K) as the coexistence of the 
high temperature cubic $F\overline{4}3m$ phase and the rhombohedral $R3m$ phase \cite{Francois1992}.  However, with 
high--resolution measurements employed here, we find that the low--temperature data cannot be fit by a combination 
of the $R3m$ phase and the $F\overline{4}3m$ phase. The middle peak shown in the lower panel of 
Figure \ref{fig1}(c), which was previously assumed to be a continuation of the cubic ($F\overline{4}3m$) phase, 
shifts significantly at the phase transition, while the higher $Q$ peak [cubic (200)] shown in the top panel 
does not split or shift at the phase transition. Because there is only one lattice parameter in the cubic phase, 
it cannot support a shift of one peak but not another, meaning that there must be another explanation for the 
additional peaks. 

We performed a search for a lower symmetry subgroup of $R3m$ (such as a monoclinic $C2/m$ phase) that could explain 
the additional peaks, but found none that did not add add extraneous peaks, not present in the data. We indexed 
the peaks that could not be explained by the $R3m$ unit cell and found they corresponded to an orthorhombic, 
polar $Imm2$ phase. $Imm2$, while a subgroup of the parent $F\overline{4}3m$ structure, is not a subgroup of 
the $R3m$ space group. The transition pathways from the $F\overline{4}3m$ phase to the $R3m$ and $Imm2$ phases 
are associated with two different order parameter directions, but with the same irreducible representation 
($T_2$) as shown in Figure \ref{fig1}(c), indicating the two subgroups are symmetrically similar.  This similarity can be seen in the diffraction data; both low temperature space groups have diffraction patterns characterized by splitting of the same cubic peaks, but the new peaks are distinct in number and position between the two phases. The difference between the two structures is most obvious in the distortion of the Mo$_4$ tetrahedron through the phase transition, indicated by the arrows in Figure \ref{fig1}(d), which leads to 2 sets of Mo$-$Mo bond lengths in the $R3m$ phase and 3 sets in the $Imm2$ phase. 
 
\begin{figure}[h]
\includegraphics[width=\linewidth]{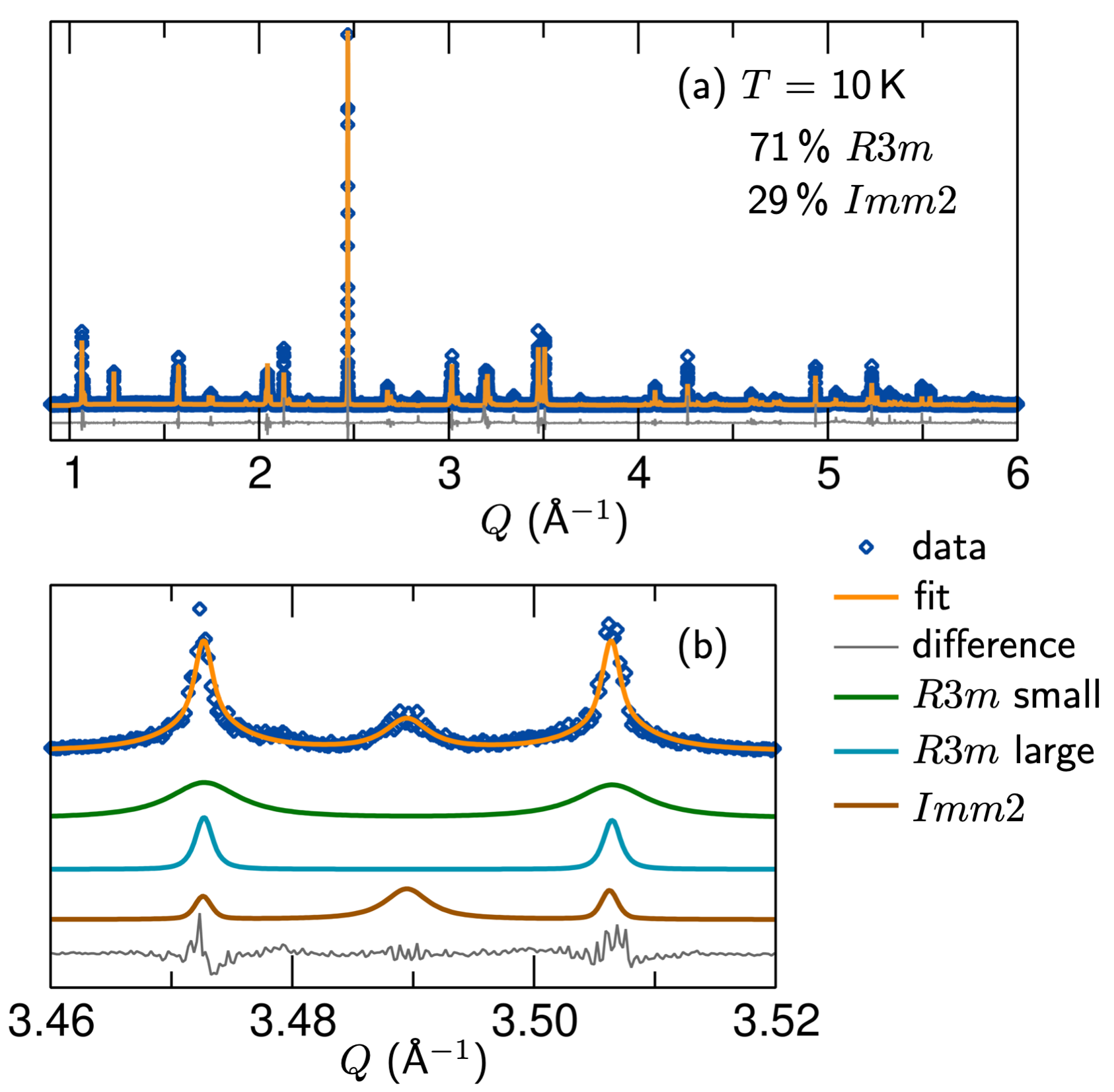}%
\caption{(a) At 10\,K, there is about 70\% of the rhombohedral phase and 30\% of the orthorhombic phase. 
(b) The complex peak shapes can be fit with two $R3m$ phases with different particle sizes, and an $Imm2$ phase with 
$hkl$-dependent strain broadening.}
\label{fig2}
\end{figure}

It is clear from the diffraction data that the $R3m$ and $Imm2$ phases form simultaneously and coexist in the 
powder. However, the structures are noticeably strained, as evidenced by the broad line shapes, and have a 
large amount of peak overlap due to their similarity, which introduces uncertainty in the determination of 
relative weight percentages of the two phases. As seen in Figure \ref{fig2}(a), our best fit to the data at 
$T$\,=\,10\,K yields about 30\% of the orthorhombic phase and 70\% of the rhombohedral phase. In order to fit 
the unusual peak shapes, we use two phases for the $R3m$ structure, one with small particles (strong effects 
of size broadening) and one with large particles, both with strain effects. The $Imm2$ phase is fit with 
$hkl$-dependent strain broadening using the Stephens peak-shape. This combination of phases to describe a 
single peak is shown in Figure \ref{fig2}(b).  At $T$\,=\,50\,K, the diffraction data is best fit by a combination 
of the cubic phase and both the $R3m$ and $Imm2$ phases, indicating both phases form simultaneously at the 
phase transition temperature and there is no phase region with only one of the two low temperature structures.  
A similar phase transition from a high symmetry structure into two subgroups has been reported in the oxide 
spinels MgCr$_2$O$_4$ and ZnCr$_2$O$_4$ and was attributed to internal strain creating a metastable 
and stable phase \cite{Kemei2013}. 
 
\begin{figure}[h]
\includegraphics[width=\linewidth]{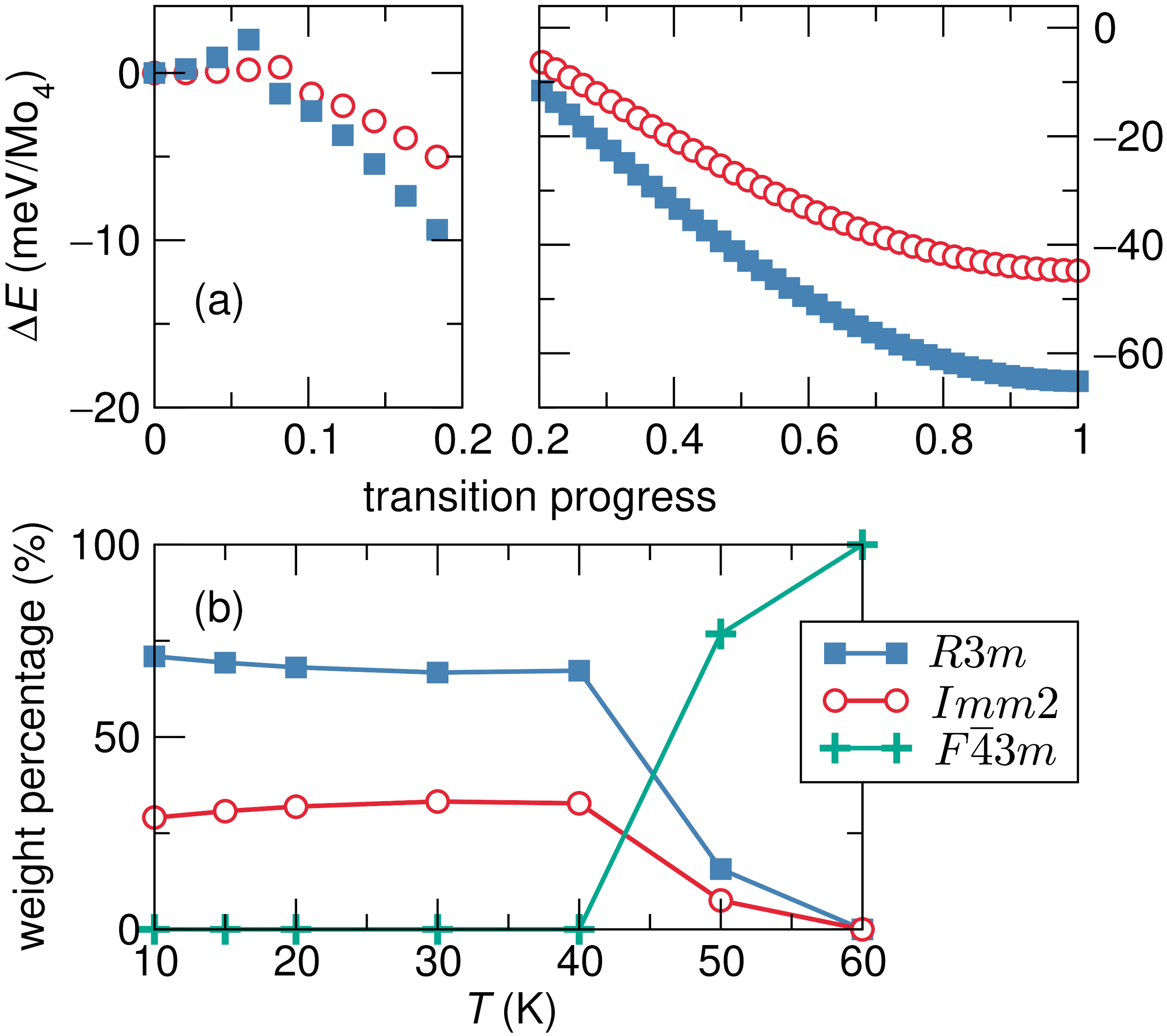}%
\caption{(a) DFT energies along the transition pathways from the cubic phase to the rhombohedral and orthorhombic 
phases. The $R3m$ structure has lower energy, making it the ground state structure, but it has a higher formation energy 
barrier, highlighted in the left panel. The energy difference is 20\,meV/Mo$_4$ cluster. (b) The phase evolution 
of GaMo$_4$Se$_8$. Upon cooling, the percentage of the $R3m$ phase slightly increases, indicating it is likely the 
ground state crystal structure.}
\label{fig3}
\end{figure}

To probe the energy landscape for these two phases, we turn to density functional theory (DFT) calculations.  
We find that while the $R3m$ phase is the ground state structure, its formation may be kinetically suppressed 
relative to the $Imm2$ phase. Using ISODISTORT, we generated distorted structures along the path from the 
$F\overline{4}3m$ phase to the $R3m$ phase and the $Imm2$ phase.  These calculations show that the orthorhombic 
structure is less stable than the rhombohedral structure by about 20\,meV/Mo$_4$ cluster, as shown in 
Figure \ref{fig3}(a).  However, the transition pathway leading to the $R3m$ phase proceeds over an energy barrier, 
while the formation of the $Imm2$ phase has a smaller barrier. Furthermore, we observe experimentally that 
the percentage of the $R3m$ phase increases with decreasing temperature, as can be seen in Figure \ref{fig3}(b). 
Thus, both computation and experiments indicate that the $R3m$ arrangement is the likely ground state. 
Nonetheless, it is possible that due to the low temperature of the phase transition ($T$\,=\,51\,K), the material 
could get trapped in the orthorhombic structure and lack the energy to convert to the rhombohedral structure.  
It is also likely that strain effects impact the energy landscape of the two phases and could stabilize the 
$Imm2$ phase.  

\begin{figure}[h!]
\includegraphics[width=\linewidth]{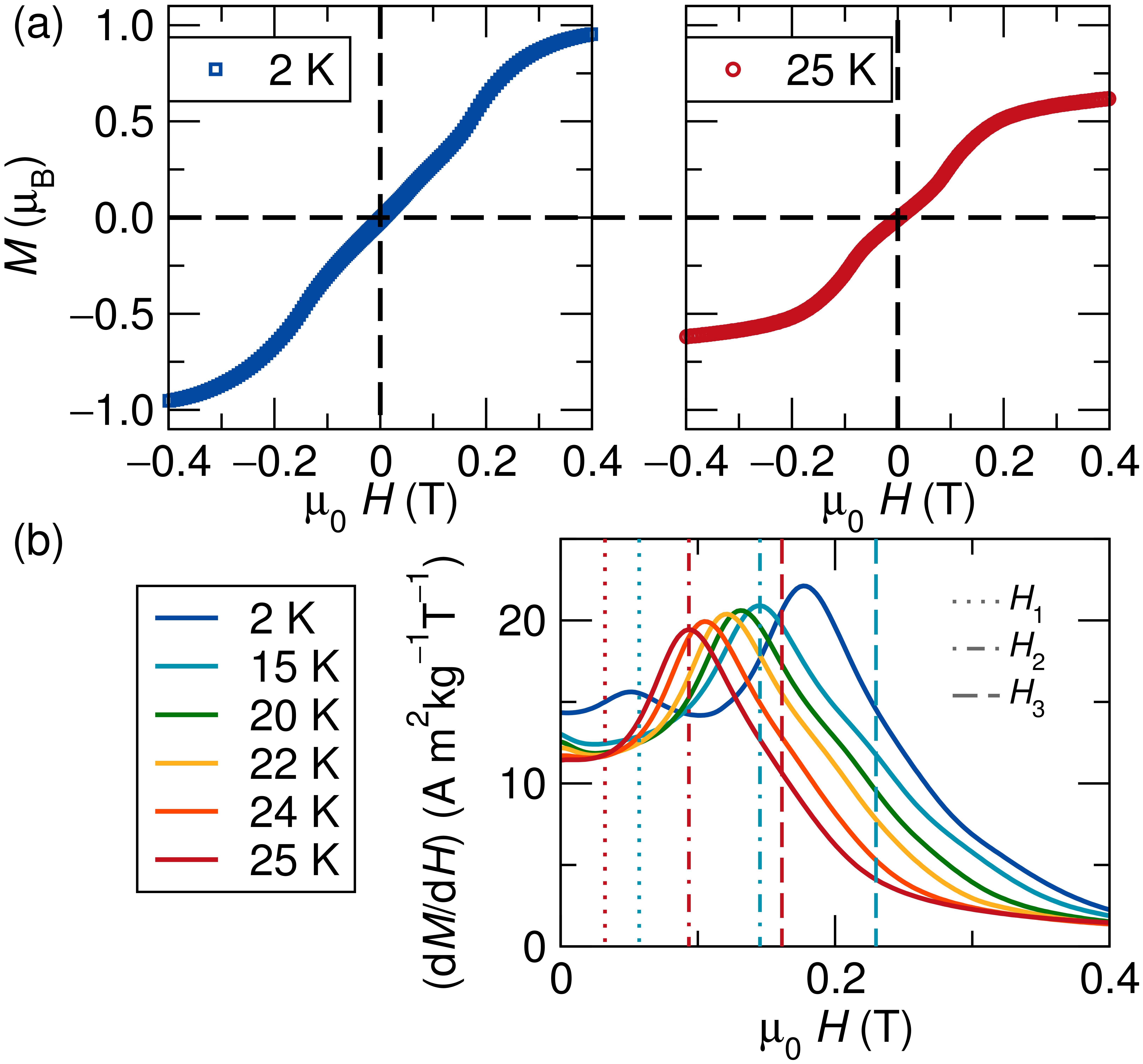}%
\caption{(a) At $T$\,=2\,K, the saturation magnetization is 1\,$\mu_B$/Mo$_4$ cluster, in agreement with DFT and 
molecular orbital theory. Multiple magnetic phase transitions are evident in field-dependent magnetization measurements. (b) Phase transitions are shown more clearly in the derivative of magnetization with respect to field. These transitions persist down to $T$\,=\,2\,K, the lowest temperature measured.}
\label{fig5}
\end{figure}
 
Analogous to other lacunar spinels, the magnetic behavior of GaMo$_4$Se$_8$ is closely tied to its crystal 
structure. Temperature--dependent magnetization data reveals largely ferromagnetic ordering with an ordering 
temperature of approximately $T$\,=\,27.5\,K, in agreement with previous reports \cite{Rastogi1987}. 
GaMo$_4$Se$_8$ has a saturation magnetization of close to 1\,$\mu_B$/Mo$_4$ cluster, as shown in Figure \ref{fig5}(a), 
in agreement with literature \cite{Rastogi1987} and molecular orbital theory. However, field--dependent magnetization measurements show the presence of multiple magnetic phase transitions at all temperatures below the ordering temperature down to $T$\,=\,2\,K, as shown in Figure \ref{fig5}(b). Resolving the nature of these phase transitions is complicated by the fact that by analogy to V--based lacunar spinels, it is likely that the magnetic phase diagram of GaMo$_4$Se$_8$ depends strongly on the direction at which the magnetic field is oriented relative to the polar axis of the crystal. As all of our data is collected on a powder sample under uniaxial applied magnetic field, different grains can have different magnetic phases, complicating the interpretation of bulk magnetization data. Nonetheless, by combining a variety of experimental and computational signatures of magnetic phase stability, we are able to understand the observed magnetic phase transitions and confirm the existence of a skyrmion phase.

\begin{figure}[h]
\includegraphics[width=\linewidth]{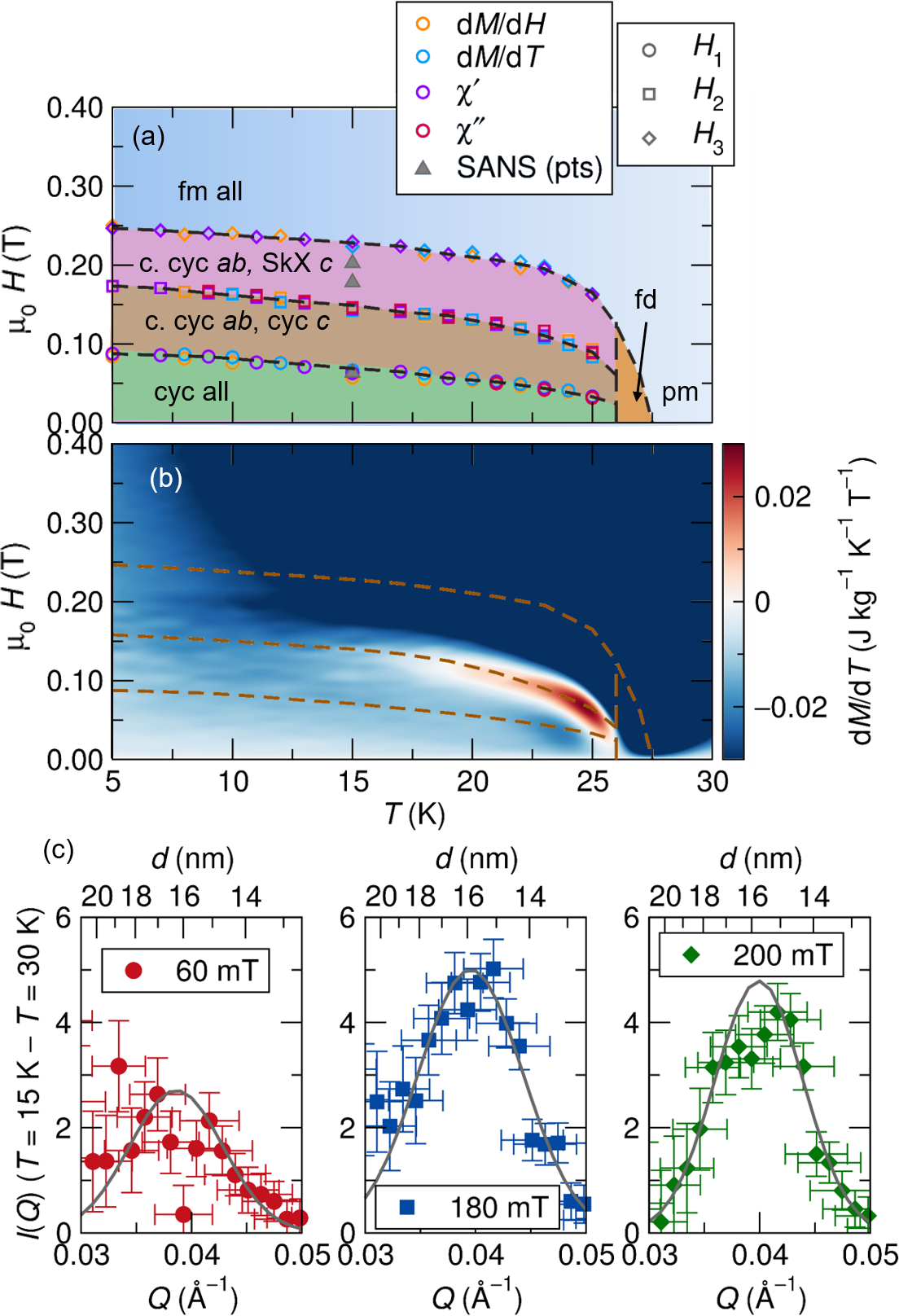}%
\caption{(a) Using multiple bulk magnetization measurement techniques, we suggest a field \textit{vs.}
temperature phase diagram for GaMo$_4$Se$_8$. Based on theoretical calculations and the phase diagrams for the 
related V--compounds, we posit that the low field transition is a transition from cycloidal into canted cycloid or ferromagnetic 
ordering in grains (denoted $ab$) where the field is mostly perpendicular to the polar axis. Upon increasing 
field, grains (denoted $c$) where the field is mostly aligned with the polar axis transition into a skyrmion 
phase and then become field polarized. Near the Curie temperature there is a fluctuation-disordered (fd) phase. 
(b) The transition into the proposed skyrmion phase in $c$-oriented grains is associated with an increase in 
magnetic entropy, shown as a red feature near the Curie temperature that broadens upon cooling. 
(c) SANS data shows the presence of approximately 16\,nm magnetic periodicity at multiple applied fields at 
$T$\,=\,15\,K. The red points are in the cycloidal phase and the blue and green are in the skyrmion phase. 
The gray lines are Gaussian fits to guide the eye.}
 \label{fig6}
 \end{figure}
 
A recently presented experimental technique for resolving magnetic phase diagrams is magnetoentropic mapping. This 
technique \cite{Bocarsly2018}, relies on the Maxwell relation that $(\partial M/\partial T)_H = (\partial S/\partial H)_T$ to identify field driven magnetic phase transitions that are associated with a change in magnetic entropy. In particular, skyrmions form as a high entropy phase, so that the formation of skyrmions is typically associated with a positive entropy anomaly \cite{Bocarsly2018, Bocarsly2019,Kautzsch2020}. On a map of ($\partial S/\partial H)_T$, shown in Figure \ref{fig6}(b), generated from temperature-dependent-magnetization measurements at various applied fields, shown in Supplementary Figure S3 \cite{SI}, we can see a red positive entropy anomaly that starts near the Curie temperature and broadens upon cooling. Overlaying this data with peaks in DC susceptibilities $(\partial M/\partial T)$, shown in Supplementary Figure S1 \cite{SI}, and $(\partial M/\partial H)$ \textit{vs.} $H$ curves, shown in Figure \ref{fig5}(b), as well as AC susceptibility measurements, shown in Supplementary Figure S2 \cite{SI}, we observe these signatures are in agreement with the transitions from the entropy map.  From these measurements we can generate the phase diagram shown in Figure \ref{fig6}(a) which suggests that the high--entropy region corresponds to a skyrmion phase with behavior similar to that seen in GaV$_4$Se$_8$.  The vertical portion of the entropy anomaly near the Curie temperature has been observed in other skyrmion host materials \cite{Bocarsly2018} and has been posited to be a Brazovskii transition into a short-range-ordered or fluctuation-disordered (fd) magnetic phase.

Susceptibility data reveals two additional transitions not visible in the magnetoentropic map, indicating the 
presence of phase transitions not associated with a large change in magnetic entropy. To unambiguously resolve 
these phase transitions, as well as to disentangle the contributions of the $R3m$ and $Imm2$ phases in the 
powder, we turn to computational modeling. Our computational analysis relies on a magnetic cluster expansion 
Hamiltonian parameterized using density functional theory (DFT), which has been recently demonstrated to yield 
reliable magnetic phase diagrams for the lacunar spinels \cite{Kitchaev2020}. We are thus able to independently 
resolve the anisotropic magnetic phase diagrams expected in the $R3m$ and $Imm2$ phases and rationalize the 
observed magnetic behavior of the biphasic powder.

 \begin{figure*}[t!]
 \includegraphics[width=.8\textwidth]{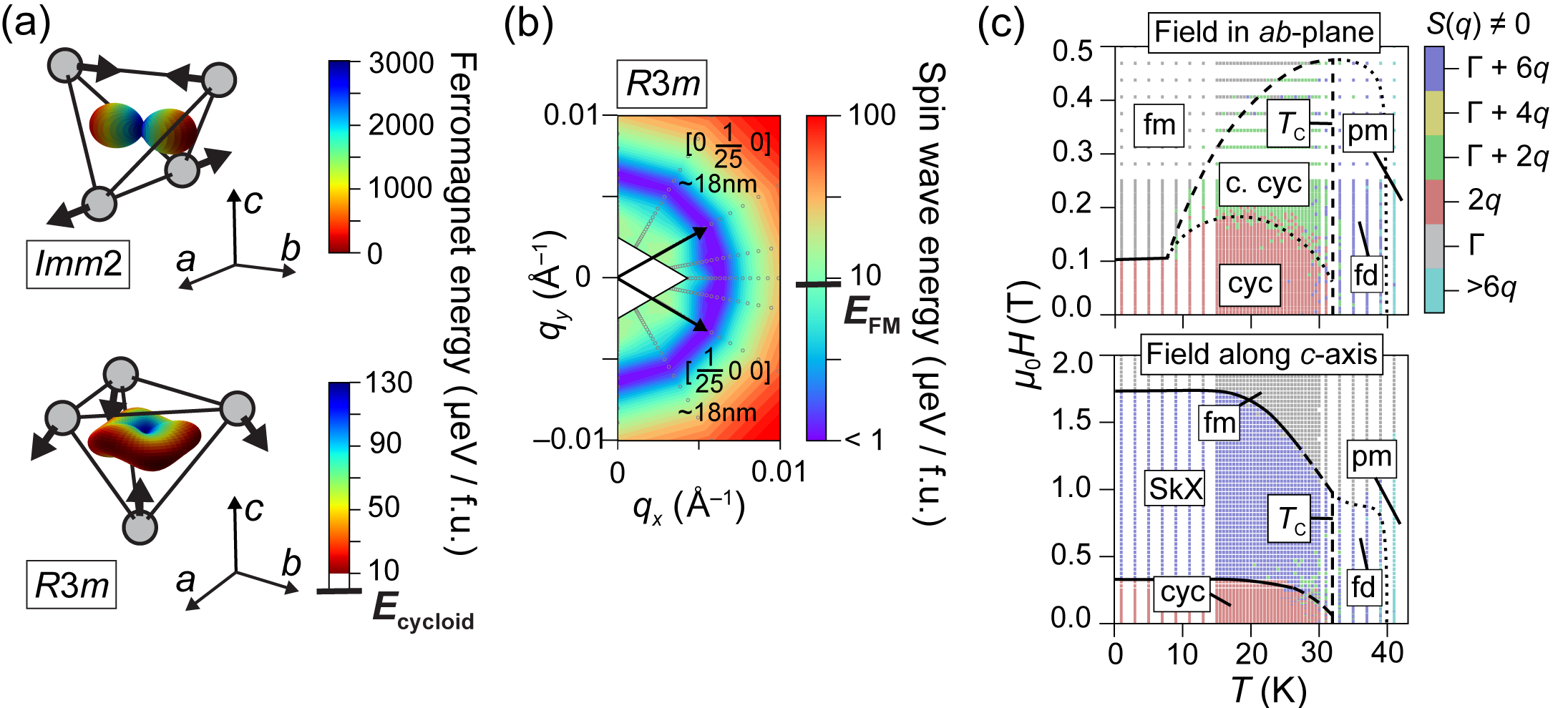}%
 \caption{(a) Computed magnetocrystalline anisotropy in the $Imm2$ and $R3m$ phases of GaMo$_4$Se$_8$. Grey circles 
 denote Mo atoms, and black arrows illustrate the direction of the Jahn-Teller distortion required to form each 
 phase. The plotted color and distance from the tetrahedron center denote the energy of a ferromagnetic spin 
 configuration oriented along the corresponding crystallographic direction. (b) Energy of spin wave configurations 
 in the ($ab$)-plane of the $R3m$ phase as a function of their $q$-vector, revealing the family of cycloid 
 ground states with a predicted wavelength of $\approx 18$ nm. (c) Computed magnetic phase diagram of the 
 $R3m$ phase for fields oriented along the $c$-axis or the ($ab$)-plane. Color denotes the magnetic structure 
 factor $S(q)$, while phase boundaries are drawn to best capture discontinuities in structure factor, 
suscetibility, heat capacity, and magnetization. Phase labels correspond to cycloid (cyc), canted cycloid 
(c. cyc), skyrmion (SkX), ferromagnet (fm), fluctuation disordered (fd), and paramagnet (pm) regions. Solid and dashed lines denote first and second order phase transitions respectively, while dotted lines denote a transition that is continuous within the resolution of our data.}
 \label{fig7}
\end{figure*}
 
First, we demonstrate that the zero--field, low--temperature ground state magnetic configuration of the 
$R3m$ phase is a cycloid with a predicted wavelength of 18\,nm, while the $Imm2$ phase is a uniaxial 
ferromagnet. Figure \ref{fig7}(a) maps out the energy of the ferromagnetic state in both phases relative 
to the energy of the ground state magnetic phase. The $Imm2$ phase strongly favors magnetization along the 
$b$-axis, parallel to the shortest Mo$-$Mo bond. The $R3m$ phase exhibits a predominantly easy--plane 
anisotropy, normal to the compressive distortion of the Mo$_4$ tetrahedron, with a three-fold modulation 
favoring spin orientation towards the corners of the tetrahedron. While the Dzyaloshinskii--Moriya 
interaction (DMI) in the two phases is comparable in magnitude and favors cycloidal spin textures in the 
($ab$)--planes \cite{Kitchaev2018}, the large magnetocrystalline anisotropy of the $Imm2$ phase suppresses 
any spin configuration deviating from the $b$-axis, forcing a ferromagnetic ground state. This behavior is in contrast to easy-axis skyrmion hosts such as GaV$_4$S$_8$, where the anisotropy is sufficiently weak to shift the skyrmion phase boundaries but not suppress them entirely \cite{Kitchaev2020}. In the $R3m$ phase, the DMI overcomes the anisotropy energy and leads to the formation of equilibrium spin cycloids at the low energy wavevectors shown in Figure \ref{fig7}(b).  The cycloids can form in any direction in the ($ab$)-plane, which is the plane perpendicular to the polar ($c$) axis of the $R3m$ structure. 

\begin{table}
\begin{ruledtabular}
\begin{tabular}{l c c c c}
	 				& $A$ & $D_{xy}$  & $D_{yx}$ & $K_u$  \\
	 				& (10$^{-13}$ J/m) & \multicolumn{2}{c}{(10$^{-4}$ J/m$^{2}$)} & (10$^{4}$ J/m$^{3}$) \\ \hline
$R3m$				& 3.48		& -2.35		& 2.35		& 7.34 \\
$Imm2$				& 4.15		& -5.00		& -0.682	& -182  \\
		
\end{tabular}
\end{ruledtabular} 
\caption{\label{table:mumag}Effective micromagnetic interaction parameters in the low-temperature limit for the $R3m$ 
and $Imm2$ phases of GaMo$_4$Se$_8$. See main text and Supplementary Figure S4 \cite{SI} for a definition of variables and coordinate systems.}
\end{table}

The low--temperature magnetic behavior of the $R3m$ and $Imm2$ phases can be summarized using a conventional 
micromagnetic free energy density functional with micromagnetic parameters given in Table \ref{table:mumag}, which is directly comparable to that of other skyrmion-host materials\cite{Takagi2017,Bordacs2017}:

\[ E = A \sum_{ij} m^2_{i,j} + \sum_{ijkn} D_{kn} \varepsilon_{ijk} m_i m_{j,n} + K_u m_u^2 \]

\noindent Here, $m$ is a magnetization unit vector, $A$ is the exchange stiffness, $D_{kn}$ is the Dzyaloshinkii 
coupling, and $K_u$ is the uniaxial anisotropy constant defined with respect to the $c$-axis of the $R3m$ phase, 
and the $b$-axis of the $Imm2$ phase. The indices $ijkn$ iterate over  Cartesian directions $(x,y,z)$, and 
$\varepsilon_{ijk}$ is the Levi-Civita symbol. The notation $m_{i,j}$ denotes a partial derivative 
$\partial m_i/\partial r_j$. The orientation of the $R3m$ and $Imm2$ crystals with respect to
the Cartesian axes is given in Supplementary Figure S4 \cite{SI}. A detailed description of this functional can be
found in the literature \cite{Kitchaev2018,Abert2019}.

Having established the low--temperature, zero--field magnetic behavior of the two phases comprising our 
GaMo$_4$Se$_8$ sample, we proceed to construct the full field \textit{vs.} temperature phase diagram of the material. 
We focus on the $R3m$ phase, as shown in Figure \ref{fig7}(c), because the $Imm2$ phase remains a uniaxial ferromagnet 
at all fields within our experimental range and does not contribute to the observed phase transitions. For fields 
oriented along the ($ab$)-plane, the phase diagram is defined by a transition from the cycloidal ground state to a 
field--polarized ferromagnetic state near $\mu_0 H$\,=\,0.1\,T. Below $\approx T_C / 3$, this phase transition is 
first order, while at higher temperatures it proceeds as a second order transition through a canted cycloid 
intermediate. When the field is oriented along the $c$-axis, the low--field cycloid undergoes a first--order 
transition into a skyrmion lattice phase at around $\mu_0 H$\,=\,0.3\,T, which appears to remain stable to very high 
fields, up to $\mu_0 H$\,=\,1.7\,T. Magnetic order persists up to $T_C$ = 32\,K, with a fluctuation-disordered 
Brazovskii region appearing immediately above $T_C$. While some quantitative discrepancies exist between this 
computational phase diagram and the experimental data, in particular in the range of fields for which skyrmions 
appear to be stable, the overall trends in phase behavior are informative for determining which phase transitions 
are being observed in our powder data.

The trends in phase behavior observed in the computational model confirm that the the high--entropy region 
observed in the magnetoentropic map likely corresponds to a skyrmion phase in $R3m$ grains with the $c$--axis 
nearly parallel to the applied field. Furthermore, we propose that the low field transition seen in the 
susceptibility data is from a cycloidal to a ferromagnetic state in $R3m$ grains where the magnetic field is 
mostly aligned perpendicular to the polar axis. The high field transition is most likely a transition from the 
skyrmion lattice to a ferromagnetic state in $R3m$ grains in which the magnetic field is mostly aligned along 
the polar ($c$) axis. Finally, the $Imm2$ phase contributes a ferromagnetic background, with no additional 
magnetic phase transitions. The similar magnitude of the exchange energy between the $Imm2$ and $R3m$ phases, approximated as $A$ in Table \ref{table:mumag}, suggests that their magnetic ordering temperatures should be quite similar, which explains why two magnetic ordering transitions are not seen in the experimental data.

To verify the presence of long--wavelength magnetic order in GaMo$_4$Se$_8$, we performed powder 
SANS measurements at 15\,K in the regions we predict to correspond to cycloid and skyrmion stability. The points 
at which SANS data was taken are indicated on our proposed phase diagram in Figure \ref{fig6}(a) with triangle markers. 
Due to the small moment of 1\,$\mu_B$/Mo$_4$ cluster and different grains in different magnetic phases, the 
signal is quite low. Nonetheless, after subtracting background data taken at $T$\,=\,30\,K (above the Curie 
temperature), we see one to two peaks in the SANS signal at all applied fields, shown in Figure \ref{fig6}(c), 
corresponding to a real-space magnetic periodicity of around 16\,nm, in acceptable agreement with our theoretical 
calculations for cycloid/skyrmion periodicity (18\,nm). 

\section{Discussion}

We have established that GaMo$_4$Se$_8$ undergoes a complex structural phase transition due to a relatively flat 
energy landscape in which multiple low temperature space groups have similar energies, as well as a competition 
between phases favored thermodynamically and kinetically. While the GaV$_4$(S/Se)$_8$ compounds exhibit only 
the $R3m$ structure at low temperature, GaMo$_4$Se$_8$ undergoes a structural transition from the high temperature 
$F\overline{4}3m$ phase into two coexisting polar space groups, rhombohedral $R3m$, which is the ground state, 
and orthorhombic $Imm2$, which is metastable. In the lacunar spinel family, GeV$_4$S$_8$ undergoes a Jahn-Teller 
distortion into an $Imm2$ structure, likely because it has one extra electron in the $t_2$ orbital relative to 
GaV$_4$S$_8$, stabilizing a different splitting of the $t_2$ orbital \cite{Bichler2008}. We speculate that in 
GaMo$_4$Se$_8$, the increased filling of the $t_2$ orbital relative to the V-compounds, along with enhanced 
spin--orbit coupling, creates an energy landscape with the $Imm2$ and $R3m$ structures much closer in energy than 
in other members of the family.

The competition between the stable and metastable crystal structures of GaMo$_4$Se$_8$ could be used to tune 
the multiferroic properties of GaMo$_4$Se$_8$, for instance by straining the material in such a way to stabilize 
one of the crystal structures. Our computational studies shows that the slight differences between the $Imm2$ 
and $R3m$ structures have a dramatic impact on the magnetic properties due to a large change in magnetocrystalline 
anisotropy. Thus, any strain which alters the relative stability of the two phases will in turn have a strong 
impact on magnetic behavior. Specifically, stabilization of the $Imm2$ phase leads to uniaxial ferromagnetism, 
while favoring the $R3m$ phase leads to long-wavelength magnetic order with a cycloidal ground state and 
N\'eel-type skyrmions. The strong emergent magnetostructural coupling in GaMo$_4$Se$_8$ provides a route towards 
the mechanical control of exotic magnetism.  

It is worth noting that extra diffraction peaks similar to those of the $Imm2$ phase in GaMo$_4$Se$_8$ were 
also observed in the low temperature diffraction patterns of the sulfide equivalent, GaMo$_4$S$_8$ and attributed 
to coexistence of the cubic phase \cite{Francois1990}. Thus, it is possible that GaMo$_4$S$_8$ could also have 
coexisting rhombohedral and orthorhombic low temperature structures, and exhibit similarly strong emergent 
magnetostructural coupling. Given the recent exciting magnetic discoveries in GaMo$_4$S$_8$\cite{Butykai2019}, 
the discovery of an $Imm2$ phase in GaMo$_4$Se$_8$ merits a re-examination of the crystal structure evolution of 
GaMo$_4$S$_8$ with synchrotron resolution.  

Finally, we have shown that the $R3m$ phase of GaMo$_4$Se$_8$ has a magnetic phase diagram with long-wavelength 
magnetic phases stable over a wide temperature and field range. GaMo$_4$Se$_8$ has a higher Curie temperature 
than the V-compounds or GaMo$_4$S$_8$ at $T_C$\,=\,27.5\,K, with cycloids and N\'eel-type skyrmions stable 
down to the lowest temperatures measured ($T$\,=\,2\,K). With a field applied along the polar axis, cycloids 
are stable from $\mu_0 H$\,=\,0\,T to around $\mu_0 H$\,=\,0.15\,T and skyrmions are stable up to $\mu_0 H$\,=\,0.25\,T at 
$T$\,=\,5\,K. SANS data show a magnetic periodicity of around 16\,nm, in good agreement with our computational 
analysis (18\,nm). Overall, the qualitative form of the phase diagram is in good agreement between our experimental 
results, computational analysis, and previous reports in the literature for materials of similar 
symmetry\cite{Bogdanov1994, Utkan2016, Randeria2016, Kitchaev2020}. However, there are deviations from experiment 
in the exact range of fields for which skyrmions are stable. The discrepancy in the cycloid to skyrmion transition 
is of a similar magnitude to the error in the Curie temperature and cycloid to in--plane ferromagnet transition, 
and is most likely a result of errors in the underlying DFT representation of GaMo$_4$Se$_8$ magnetism, as well 
as deviations in the $g$-factor from our assumed value of 2. The much larger discrepancy in the upper bound on 
skyrmion stability versus applied field is more likely a result of stray fields, which we neglect in our 
simulations. Finally, while we do not observe the complexity in magnetic ordering vector reported in 
GaMo$_4$S$_8$ \cite{Butykai2019}, this difference may be a consequence of the sensitivity of data we are able 
to obtain from a powder sample as compared to a single crystal.

\section{Conclusion}

A combined experimental and computational investigation of the low--temperature structural and magnetic 
phase behavior of the GaMo$_4$Se$_8$ lacunar spinel has been carried out. We show that this material undergoes 
a Jahn--Teller distortion at $T$\,=\,51\,K into two co-existing structures -- a metastable $Imm2$ configuration 
and the ground state $R3m$ phase. These two phases both order magnetically at $T_C$\,=\,27.5\,K, but exhibit 
dramatically different magnetic behavior due to differences in their magnetocrystalline anisotropy. While the 
$Imm2$ phase is found from computation to be a uniaxial ferromagnet, the $R3m$ phase favors a 16\,nm cycloid 
ground state, and a N\'eel--skyrmion lattice under applied field. We conclude that the coexistence of these 
two phases with dramatically different magnetic properties makes GaMo$_4$Se$_8$ a promising material for realizing 
mechanical control over exotic magnetism, as small energy differences separating the $Imm2$ and $R3m$ phases 
translate into a very large change in magnetic behavior.

\begin{acknowledgments}
This research was supported by the National Science Foundation (NSF) under DMREF Award DMR-1729489. 
Partial support from the NSF Materials Research Science and Engineering Center (MRSEC) 
at UC Santa Barbara, DMR-1720256 (IRG-1) is gratefully acknowledged (D. A. K. and A.V.dV.) 
Use of the Shared Experimental Facilities of the UCSB MRSEC (NSF DMR 1720256) is acknowledged. The UCSB MRSEC 
is a member of the NSF-supported Materials Research Facilities Network (www.mrfn.org).  
We also acknowledge support from the Center for Scientific Computing (NSF DMR-1720256 and NSF CNS-1725797), 
as well as the National Energy Research Scientific Computing Center, a DOE Office of Science User Facility 
supported by DOE DE-AC02-05CH11231. Use of the Advanced Photon Source at Argonne National Laboratory was 
supported by the U. S. Department of Energy, Office of Science, Office of Basic Energy Sciences, under 
Contract No. DE-AC02-06CH11357.  We thank Dr. S. Lapidus of 11-BM at the Advanced Photon Source for helpful contributions. We acknowledge the support of Dr. M. Bleuel and the National Institute of Standards and Technology, U.S. Department of Commerce, in providing the neutron research facilities used in this work. 
\end{acknowledgments}

%

\end{document}


\title{Structural evolution and skyrmionic phase diagram of the lacunar spinel GaMo$_4$Se$_8$}

\author{Emily C. Schueller}
\author{Daniil A. Kitchaev}
\author{Julia L. Zuo}
\author{Joshua D. Bocarsly}
\author{Joya A. Cooley}
\author{Anton Van der Ven}
\author{Stephen D. Wilson}
\affiliation{Department of Materials and Materials Research Laboratory, University of California, Santa Barbara, Santa Barbara, CA, 93106}
\author{Ram Seshadri}
\affiliation{Department of Materials and Materials Research Laboratory, University of California, Santa Barbara, Santa Barbara, CA, 93106}
\affiliation{Department of Chemistry and Biochemistry, University of California, Santa Barbara, Santa Barbara, CA, 93106}

\maketitle

\renewcommand{\thefigure}{S\arabic{figure}}
\pagebreak
\begin{figure}
\includegraphics[width=0.7\textwidth]{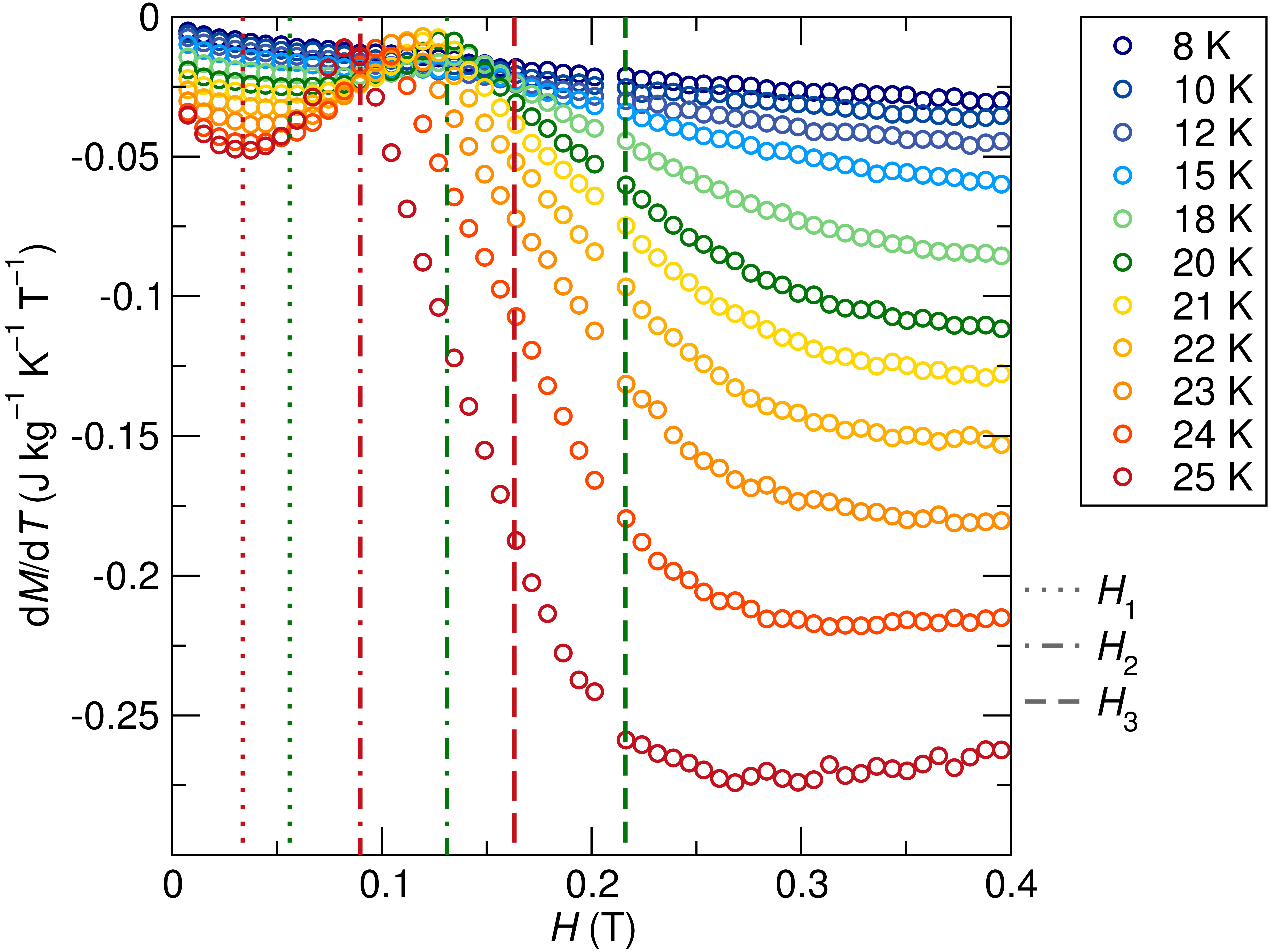}
\caption{d$M$/d$T$ vs field data at various temperatures. Some characteristic locations of the three magnetic transitions with field are shown over the data.}
\end{figure}

\begin{figure}
\includegraphics[width=0.7\textwidth]{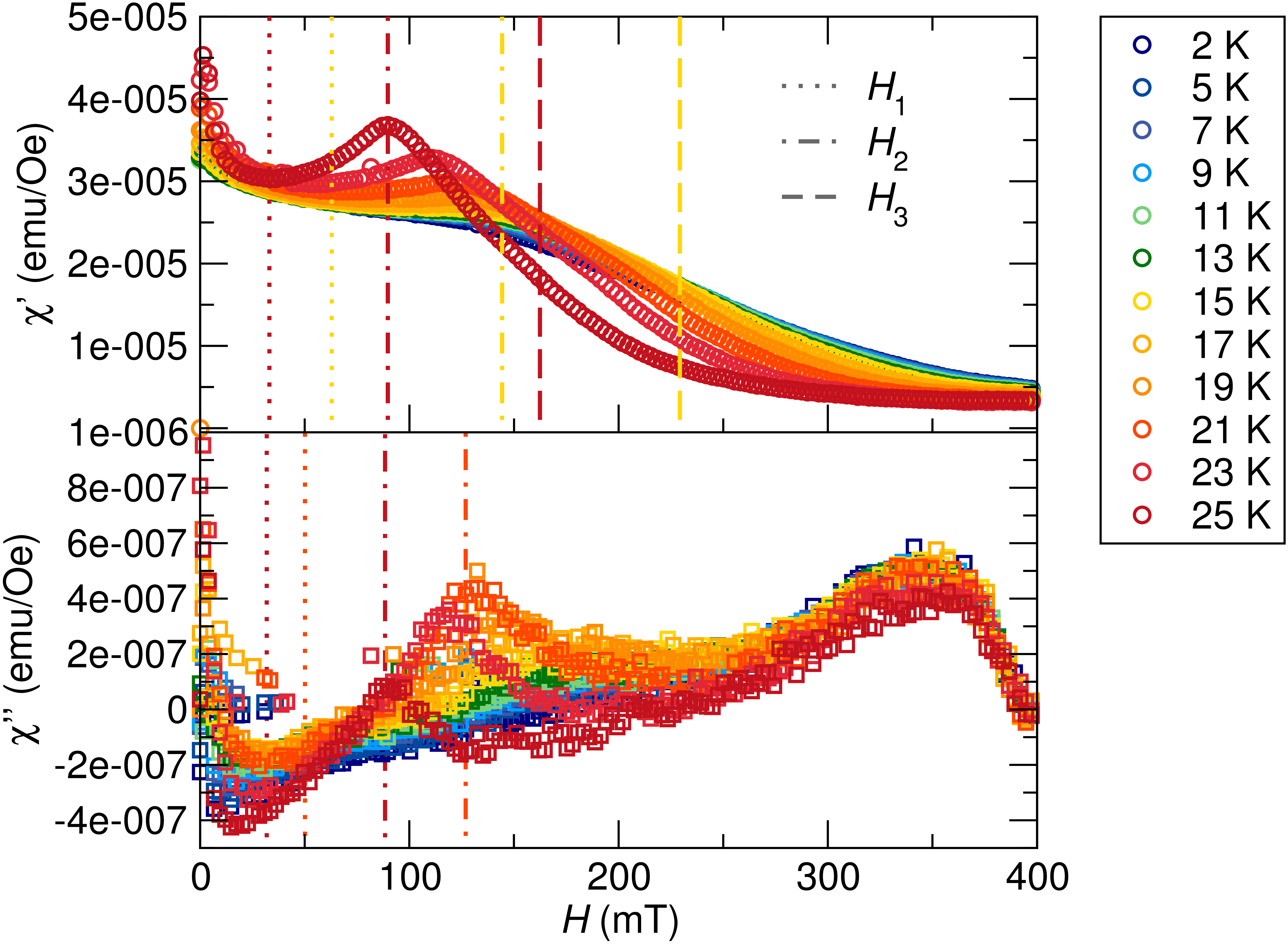}
\caption{AC $\chi$' and $\chi$'' data at various temperatures. Some characteristic locations of the three magnetic transitions with field are shown over the data. Only the first two transitions can be elucidated from the $\chi$'' data.}
\end{figure}

\begin{figure}
\includegraphics[width=0.5\textwidth]{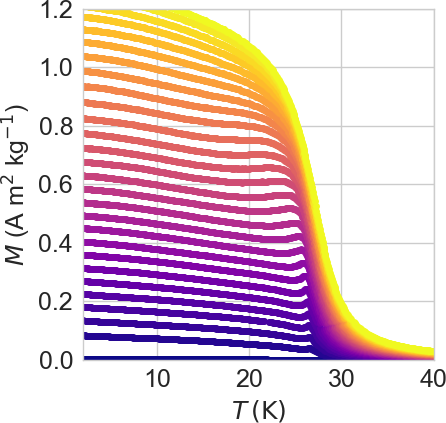}
\caption{Magnetization vs temperature data at various applied fields. This data was used to generate the magnetoentropic map by taking the derivative of magnetization with respect to temperature.}
\end{figure}

\begin{figure}
\includegraphics[width=0.9\textwidth]{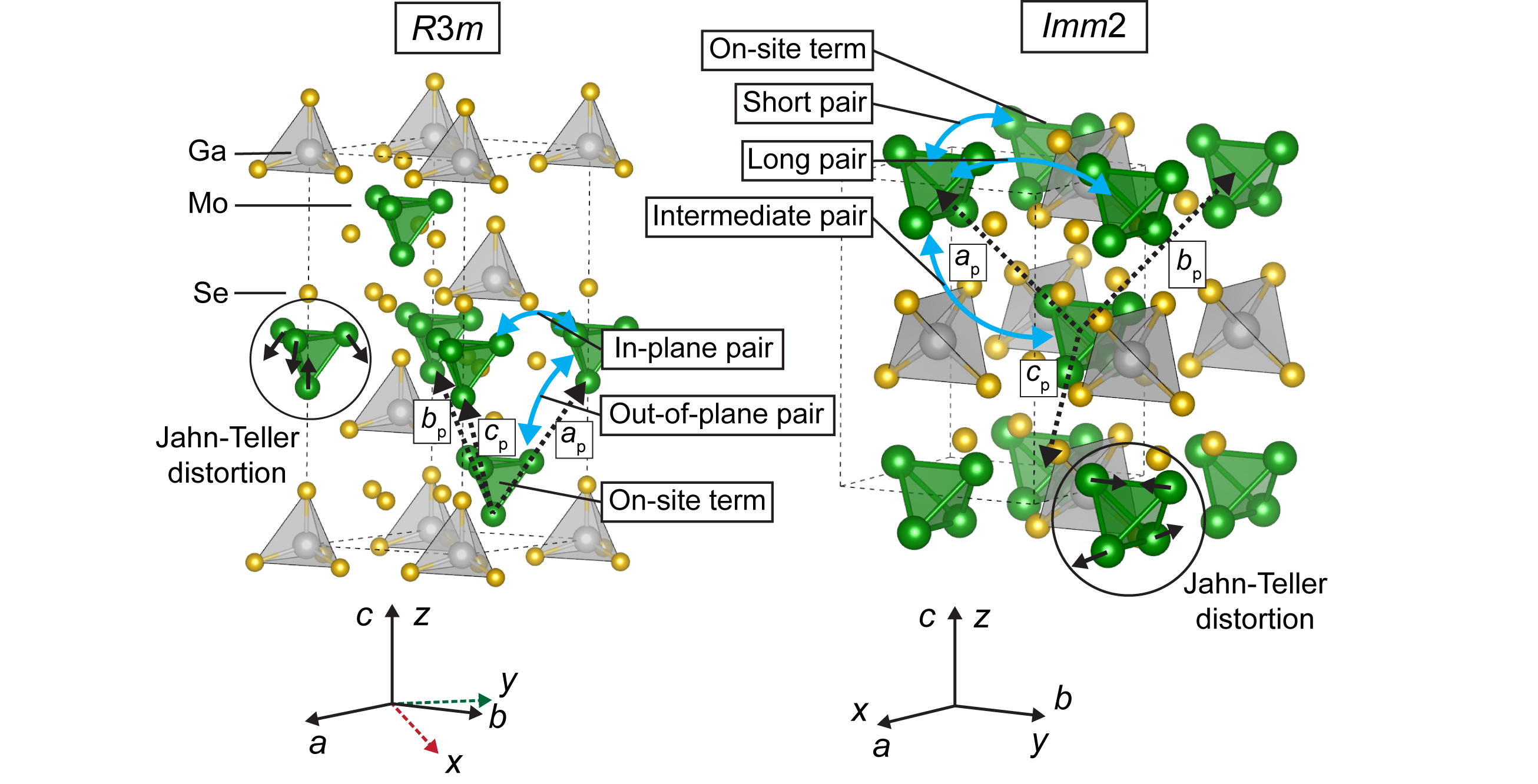}
\caption{Structure of the $R3m$ and $Imm2$ phases of GaMo${}_4$Se$_8$ in their conventional unit cells, along with the primitive cell axes ($a_{\text{p}}$, $b_{\text{p}}$, $c_{\text{p}}$). The cartesian coordinate system for each phase further defines the axes used to define the magnetic Hamiltonian. Finally, the symmetrically distinct on-site and pair interactions appearing in the magnetic Hamiltonian of each phase are shown.}
\end{figure}

\begin{figure}
\includegraphics[width=0.9\textwidth]{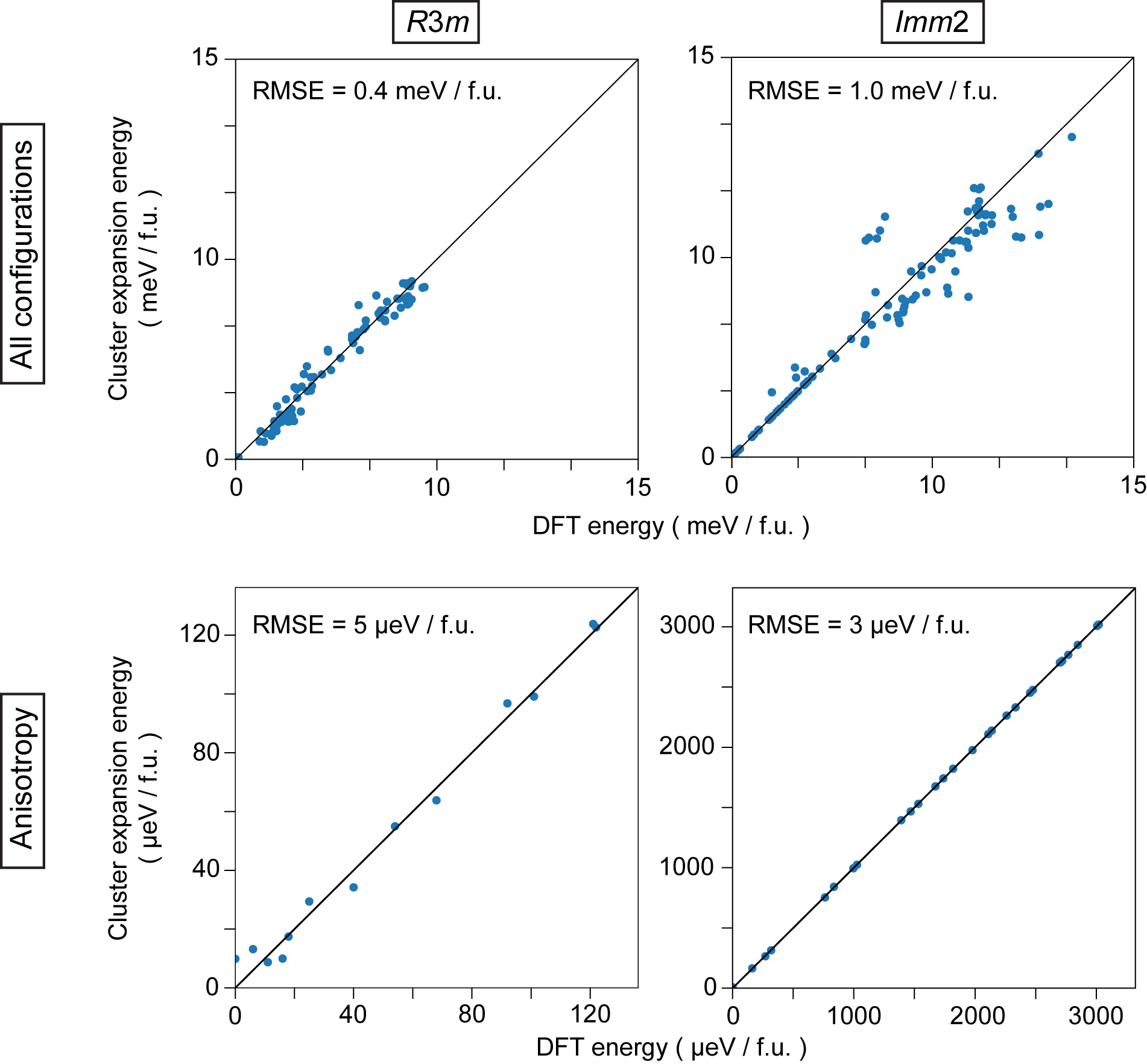}
\caption{Magnetic cluster expansion fitting error for the $R3m$ and $Imm2$ phases of GaMo${}_4$Se$_8$. Each point denotes the energy of a symmetrically distinct spin configuration, as computed using DFT and the fitted cluster expansion Hamiltonian. The magnetocrystalline anisotropy terms are fitted independently to account for the difference in energy scale between these terms and spin-spin interactions, as described in the main text.}
\end{figure}

\begin{table}
\begin{ruledtabular}
\begin{tabular}{l r l c}
Cluster type & \multicolumn{2}{c}{Basis function}  & $J$(meV) \\ \hline
On-site		& $\phi^{A}_1$	& $\sqrt{2} |2,0\rangle$ & 0.027(1) \\
$r = (0, 0, 0)$	& $\phi_2^{A}$		& $-i \left(|4,-3\rangle + |4,3\rangle\right) $	& -0.008(1)\\
(1 equiv.)			& $\phi^{A}_3$		& $\sqrt{2} |4,0\rangle$ & -0.006(1) \\ \hline
Out-of-plane& $\phi^{E}_7$		& $\frac{\sqrt{2}}{3} |1,1;0,0\rangle$	& 1.38(6) \\
$r_1 = (0, 0, 0)$	& $\phi^{D}_8$		& $\frac{i}{3} \left( |1,1;1,1\rangle - |1,1;1,-1\rangle \right)$ & -0.06(9) \\
$r_2 = (-1, 0, 0)$	& $\phi^{A}_9$		& $-\frac{i}{3} \left( |1,1;2,1\rangle + |1,1;2,-1\rangle \right)$ & 0.1(2)  \\
(3 equiv.)			& $\phi^{A}_{10}$	& $\frac{\sqrt{2}}{3} |1,1;2,0\rangle $ & 0 \\
			& $\phi^{A}_{11}$	& $\frac{1}{3} \left( |1,1;2,2\rangle + |1,1;2,-2\rangle \right)$ & 0.6(3) \\  \hline
In-plane	& $\phi^{E}_{12}$	& $\frac{\sqrt{2}}{3} |1,1;0,0\rangle$ & 1.06(6) \\
$r_1 = (0,0,0)$	& $\phi^{D}_{13}$	& $(\frac{1}{6} + \frac{i}{\sqrt{12}})|1,1;1,1\rangle +	$ & 0.3(1) \\
$r_2 = (-1,0,1)$		&				& $(\frac{1}{6} - \frac{i}{\sqrt{12}})|1,1;1,-1\rangle$ &  \\
(3 equiv.)			& $\phi^{D}_{14}$	& $\frac{-i\sqrt{2}}{3} |1,1;1,0\rangle$ & 0 \\
			& $\phi^{A}_{15}$	& $\frac{\sqrt{2}}{3} |1,1;2,0\rangle$ & 0 \\
			& $\phi^{A}_{16}$	& $(\frac{1}{\sqrt{12}}-\frac{i}{6})|1,1;2,1\rangle - $ & 0 \\
			& 				& $(\frac{1}{\sqrt{12}}+\frac{i}{6})|1,1;2,-1\rangle$ & \\
			& $\phi^{A}_{17}$	& $(\frac{1}{6} - \frac{i}{\sqrt{12}})|1,1;2,2\rangle +	$ & 0 \\
			&				& $(\frac{1}{6} + \frac{i}{\sqrt{12}})|1,1;2,-2\rangle$ & \\

\end{tabular}
\end{ruledtabular}
\caption{\label{table:hamiltonian} Magnetic Hamiltonian for the $R3m$ phase of GaMo$_4$Se$_8$, consisting of basis functions $\phi$ and parametrized interaction strengths $J$. Cluster site coordinates and basis functions are given for the reference cluster, in lattice coordinates with respect to the primitive lattice vectors ($a_{\text{p}}$, $b_{\text{p}}$, $c_{\text{p}}$) given in Supplementary Figure 3. The number of equivalents for each cluster refers to the number of symmetrically--equivalent clusters of this type per primitive cell. Basis functions are defined in terms of spherical harmonics $|l,m \rangle = \sqrt{4\pi} Y^{l}_{m}(\phi, \theta)$ for the on-site terms and Clebsch--Gordan functions $|l_1,l_2;L,M \rangle = 4\pi \sum_{m_1,m_2} c^{l_1,l_2,L}_{m_1,m_2,M}Y^{l_1}_{m_1}(\phi_1, \theta_1)Y^{l_2}_{m_2}(\phi_2, \theta_2)$ for pair clusters $(r_1,r_2)$. Spin angles $(\phi, \theta)$ are given in spherical coordinates with respect to the global coordinate system defined in Supplementary Figure 4. Basis function superscripts denote whether the interaction corresponds to exchange (E), DMI (D), or anisotropy (A). Parenthesis in the $J$ vector components denote uncertainty in the last digit.}
\end{table}

\begin{table}
\begin{ruledtabular}
\begin{tabular}{l r l c}
Cluster type & \multicolumn{2}{c}{Basis function}  & $J$(meV) \\ \hline
On-site				& $\phi^{A}_1$		& $\sqrt{2} |2,0\rangle$ 													& 0.313(1) \\
$r = (0, 0, 0)$		& $\phi_2^{A}$		& $|2,2\rangle + |2,-2\rangle$ 												& 0.551(1)\\
(1 equiv.)			& $\phi^{A}_3$		& $|4,4\rangle + |4,-4\rangle$ 												& -0.003(1) \\
					& $\phi^{A}_4$		& $|4,2\rangle + |4,-2\rangle$												& -0.002(1) \\
					& $\phi^{A}_5$		& $\sqrt{2} |4,0\rangle$ 													& 0 \\ \hline
Short 				& $\phi^{E}_{10}$	& $\sqrt{2} |1,1;0,0\rangle$												& 0.93(7) \\
$r_1 = (0, 0, 0)$	& $\phi^{D}_{11}$	& $|1,1;1,1\rangle + |1,1;1,-1\rangle $ 									& -0.18(9) \\
$r_2 = (0, -1, -1)$	& $\phi^{A}_{12}$	& $|1,1;2,-2\rangle + |1,1;2,-2\rangle $ 									& -0.2(2)  \\
(1 equiv.)			& $\phi^{A}_{13}$	& $\sqrt{2} |1,1;2,0\rangle $ 												& -0.2(1) \\ \hline
Intermediate		& $\phi^{E}_{14}$	& $\frac{\sqrt{2}}{4} |1,1;0,0\rangle$										& 0.2(1) \\
$r_1 = (1,0,0)$		& $\phi^{D}_{15}$	& $\frac{1}{4}\left(|1,1;1,1\rangle + |1,1;1,-1\rangle\right)$				& 0.3(2) \\
$r_2 = (0,-1,-1)$	& $\phi^{D}_{16}$	& $\frac{i}{4}\left(|1,1;1,1\rangle - |1,1;1,-1\rangle\right)$				& -0.1(1) \\
(4 equiv.)			& $\phi^{D}_{17}$	& $\frac{-i\sqrt{2}}{4} |1,1;1,0\rangle$									& -0.3(2) \\
					& $\phi^{A}_{18}$	& $\frac{1}{4}\left(|1,1;2,2\rangle + |1,1;2,-2\rangle\right)$				& 0.2(3)\\
					& $\phi^{A}_{19}$	& $\frac{i}{4}\left(|1,1;2,2\rangle - |1,1;2,-2\rangle\right)$				& -0.1(4) \\
					& $\phi^{A}_{20}$	& $\frac{-1}{4}\left(|1,1;2,1\rangle - |1,1;2,-1\rangle\right)$	 			& 0.4(5) \\
					& $\phi^{A}_{21}$	& $\frac{-i}{4}\left(|1,1;2,1\rangle + |1,1;2,-1\rangle\right)$ 			& 0.5(5) \\
					& $\phi^{A}_{22}$	& $\frac{\sqrt{2}}{4} |1,1;2,0\rangle$				 						& 0.1(1) \\ \hline
Long 				& $\phi^{E}_{23}$	& $\sqrt{2} |1,1;0,0\rangle$												& 0.76(8) \\
$r_1 = (1, 0, 0)$	& $\phi^{D}_{24}$	& $i\left( |1,1;1,-1\rangle - |1,1;1,1\rangle \right)$ 						& 0.24(9) \\
$r_2 = (0, 0, -1)$	& $\phi^{A}_{25}$	& $|1,1;2,2\rangle + |1,1;2,-2\rangle $ 									& -0.1(1) \\
(1 equiv.)			& $\phi^{A}_{26}$	& $\sqrt{2} |1,1;2,0\rangle $ 												& 0.1(1) \\ 
\end{tabular}
\end{ruledtabular}
\caption{\label{table:hamiltonian} Magnetic Hamiltonian for the $Imm2$ phase of GaMo$_4$Se$_8$, following the format given in the caption of Supplementary Table 1.}
\end{table}